\documentclass[prl,showpacs,floatfix,twocolumn]{revtex4-1}

\usepackage{graphicx}
\usepackage{dcolumn}
\usepackage{float}
\usepackage{amsmath}

\newcommand{\comment}[1]{}

\begin{document}

\title{\boldmath  High energy fluctuation spectra in cuprates from infrared optical spectroscopy \unboldmath}

\author{Jungseek Hwang}
\email[Corresponding author:~]{jungseek@skku.edu}
\affiliation{Department of Physics, Sungkyunkwan University, Suwon, Gyeonggi-do 440-846, Republic of Korea}
\author{J. P. Carbotte}
\affiliation{Department of Physics and Astronomy, McMaster University, Hamilton, ON L8S 4M1, Canada \\The Canadian Institute for Advanced Research, Toronto, ON M5G 1Z8 Canada}

\date{\today}

%
%
\begin{abstract}
Coupling of the charge carriers in the high temperature superconducting oxides to bosonic modes has been widely reported using a variety of experimental probes. These include angular resolved photoemission (ARPES), scanning tunnelling spectroscopy (STS), Raman scattering (RS) and infrared optical spectroscopy (IRS). The energy scale investigated has been mostly limited to a relatively small range up to 300 meV or so. Although some ARPES experiments report boson structure up to 800 meV in the dressed electron dispersion curves the data are not analyzed to recover the spectral density of the fluctuation spectrum. We have extended to higher energies up to 2.2 eV the usual maximum entropy technique used to invert optical data so as to obtain an electron-boson spectral density. This has required that we include in our inversions, the calculated (LDA) particle-hole symmetrized energy dependent electronic density of states (DOS). Our analysis reveals that significant spectral weight remains in the fluctuation spectra up to 2.2 eV in the Bi-2212 family and to 1.2 eV in Bi-2201 for all doping levels considered.

\end{abstract}


\maketitle
%

\section{I. Introduction}

The excitation spectrum to which the charge carriers in the high T$_c$ cuprates are coupled is of great interest and has been studied extensively using many different experimental probes\cite{carbotte:2011}. The techniques that have been brought to bare on the problem include infrared optical spectroscopy (IRS)\cite{carbotte:2011,yang:2009a,hwang:2007,heumen:2009,yang:2009,hwang:2008c}, angular resolved photoemission (ARPES)\cite{carbotte:2011,garcia:2010,shen:1995,damascelli:2003,johnson:2001,shi:2004,zhang:2008,bok:2010,schachinger:2008}, Raman scattering\cite{muschuler:2010} as well as scanning and other tunnelling spectroscopy (STS)\cite{zasadzinski:2006,lee:2006,pasupathy:2008,jenkins:2009}. Each of these various techniques has special advantages and also drawbacks. For example fine momentum selectivity can be achieved in ARPES. While this is obviously very desirable for some purposes, a momentum average over all charge carriers may be preferred in other cases. Also, ARPES is a surface probe\cite{garcia:2010,damascelli:2003} while one may well want bulk information. STS is likewise a surface probe and is local in space\cite{lee:2006,pasupathy:2008,jenkins:2009}, but a spacial average of STS data is routinely done. On the other hand optical absorption and Raman scattering are bulk probes. Optics gives directly a momentum average\cite{carbotte:2011}, while Raman\cite{muschuler:2010} in its B$_{2g}$, B$_{1g}$ polarization modes samples preferentially only a select set of charge carriers around nodal and antinodal direction respectively. Despite these differences a great deal of agreement exists between the excitation spectra recovered from the various set of data\cite{carbotte:2011}.

It is usually assumed that the structure seen in the data can be described within a boson exchange mechanism\cite{carbotte:2011,marsiglio:1998,schachinger:2006,sharapov:2005}. This allows one to extract an electron-boson spectral density denoted by $I^2\chi(\omega)$ which describes the excitations involved in the inelastic scattering. The shape and size of the spectral density carries information on these excitations. While infrared data has been extensively employed in such studies\cite{carbotte:2011}, the analysis has generally been limited to the lower energy range of the spectra up to 300 meV or so. One reason for this has been the assumption, in most analysis of data, that the electronic density of states is constant over the range of energies of interest. On the other hand there has been suggestions from ARPES data that boson structure may persist up to 800 meV. Markiewicz {\it et al.}\cite{markiewicz:2012} have also reviewed a number of theoretical studies that suggest that the excitation spectrum in the optimally doped cuprates may extend to high energies into range associated with the incoherent part of the bands. In this paper we present new inversions of optical data with the aim of recovering an electron-boson spectral density up to 2.0 eV or even beyond. On this larger energy scale it is no longer clear that the band structure density of states can be considered to be constant. Thus it is necessary to account for this possibility in the inversion process\cite{sharapov:2005,mitrovic:1983,mitrovic:1983:2,hwang:2013,hwang:2012,hwang:2006}. We still employ a maximum entropy technique to go from the data on the optical scattering rate to an estimate of $I^2\chi(\omega)$ but we now include explicitly the energy dependence of the LDA band structure density of states $N(\omega)$. The paper is structured as follows. In section II we review the relationship between the conductivity and the spectral density. The formalism is based on a Kubo formula for the conductivity in a boson exchange model, and on simplifications which allow for easier inversion of data. We also give a brief introduction to the maximum entropy inversion technique which we use. Results are presented in section III. A summary and conclusions are found in section IV.

\section{II. Formalism}

In terms of the frequency and temperature dependent optical self energy $\Sigma^{op}(T,\omega)$ the optical conductivity $\sigma(T,\omega)$ takes the form\cite{hwang:2004,hwang:2013,hwang:2012}
\begin{equation}\label{eq1}
\sigma(T,\omega) = \frac{i}{4\pi}\frac{\Omega_p^2}{\omega-2\Sigma^{op}(T,\omega)}
\end{equation}
where $\Omega_p$ is the plasma energy. The imaginary part of $-2\Sigma^{op}(T,\omega)$ defines an optical scattering rate $1/\tau^{op}(T,\omega)$ and the real part a renormalized optical effective mass $m^{*op}(T,\omega)/m$ with $\omega[m^{*op}(T,\omega)/m-1] = -2Re\Sigma^{op}(T,\omega)$. The optical mass enhancement $\lambda^{op}(T,\omega)$ is defined as $1+\lambda^{op}(T,\omega) = m^{*op}(T,\omega)/m$. It is the energy and temperature dependence of these functions which carry the information on the inelastic scattering here assumed to be due to coupling to bosons. In conventional superconductors it is this inelastic scattering that leads to the so called strong coupling corrections\cite{carbotte:1995,carbotte:1986,mitrovic:1980} to conventional BCS theory. Additional features can also provide further corrections such as momentum anisotropies\cite{leavens:1971,odonovan:1995b,odonovan:1995a}. In general $\sigma(T,\omega)$ of Eqn. (\ref{eq1}) can be calculated from a Kubo formula which involves an average over all momentum which makes this quantity less sensitive to anisotropies and provides an average measure of the coupling of charge carriers to the bosons. For a boson exchange model with electron-boson spectral density denoted by $I^2\chi(\omega)$, Allen\cite{allen:1971} derived a very simple approximate, but analytic formula for the connection between the optical scattering rate at zero temperature and the spectral density of states $I^2\chi(\omega)$ which turns out to be most convenient for analysis of the optical data. Allen's derivation involved ordinary perturbation theory and was for zero temperature ($T =$ 0). Since then it has been derived rigorously from the usual Kubo formula and generalized to finite temperature\cite{shulga:1991}. The case when important energy dependence exists in the electronic density of states $N(\omega)$ was further considered by Mitrovic and Fiorucci \cite{mitrovic:1985} at $T =$ 0 and generalized to finite $T$ by Sharapov and Carbotte\cite{sharapov:2005} who start from a Kubo formula which explicitly includes energy dependence in $N(\omega)$ and introduced simplifications which lead to an integral equation which is much more convenient for the analysis of optical data. The formula of Sharapov and Carbotte\cite{sharapov:2005} is
\begin{eqnarray}\label{eq1a}
\frac{1}{\tau^{op}(T,\omega)}&=&\frac{\pi}{\omega}\int^{\infty}_{0} d\Omega I^2\chi(T,\Omega)\int^{+\infty}_{-\infty}dz \tilde{N}(z-\Omega)\times\\\nonumber &&[n_B(\Omega)+1-f(z-\Omega)][f(z-\omega)-f(z+\omega)]
\end{eqnarray}
where $n_B(\Omega)$ and $f(\Omega)$ are respectively the Bose-Einstein and Fermi-Dirac distributions at finite $T$ and $\tilde{N}(\omega)$ is the particle-hole symmetrized electron density of states $\tilde{N}(\omega) = 1/2[N(\omega)+N(-\omega)]$. This generalized formula properly reduces to that of Shulga {\it et al.}\cite{shulga:1991} when $N(z)$ is assumed constant and reduces further to Allen's form when $T =$ 0. For zero temperature but variable density of states it reduces to the formula of Mitrovic and Fiorucci\cite{mitrovic:1985},
\begin{equation}\label{eq1b}
\frac{1}{\tau^{op}(T=0,\omega)}=\frac{2\pi}{\omega}\int^{\omega}_{0} d\Omega I^2\chi(\Omega)\int^{\omega-\Omega}_{0}d\omega' \tilde{N}(\omega').
\end{equation}

\begin{figure}[t]
  \vspace*{-1.0 cm}%
  \centerline{\includegraphics[width=3.0 in]{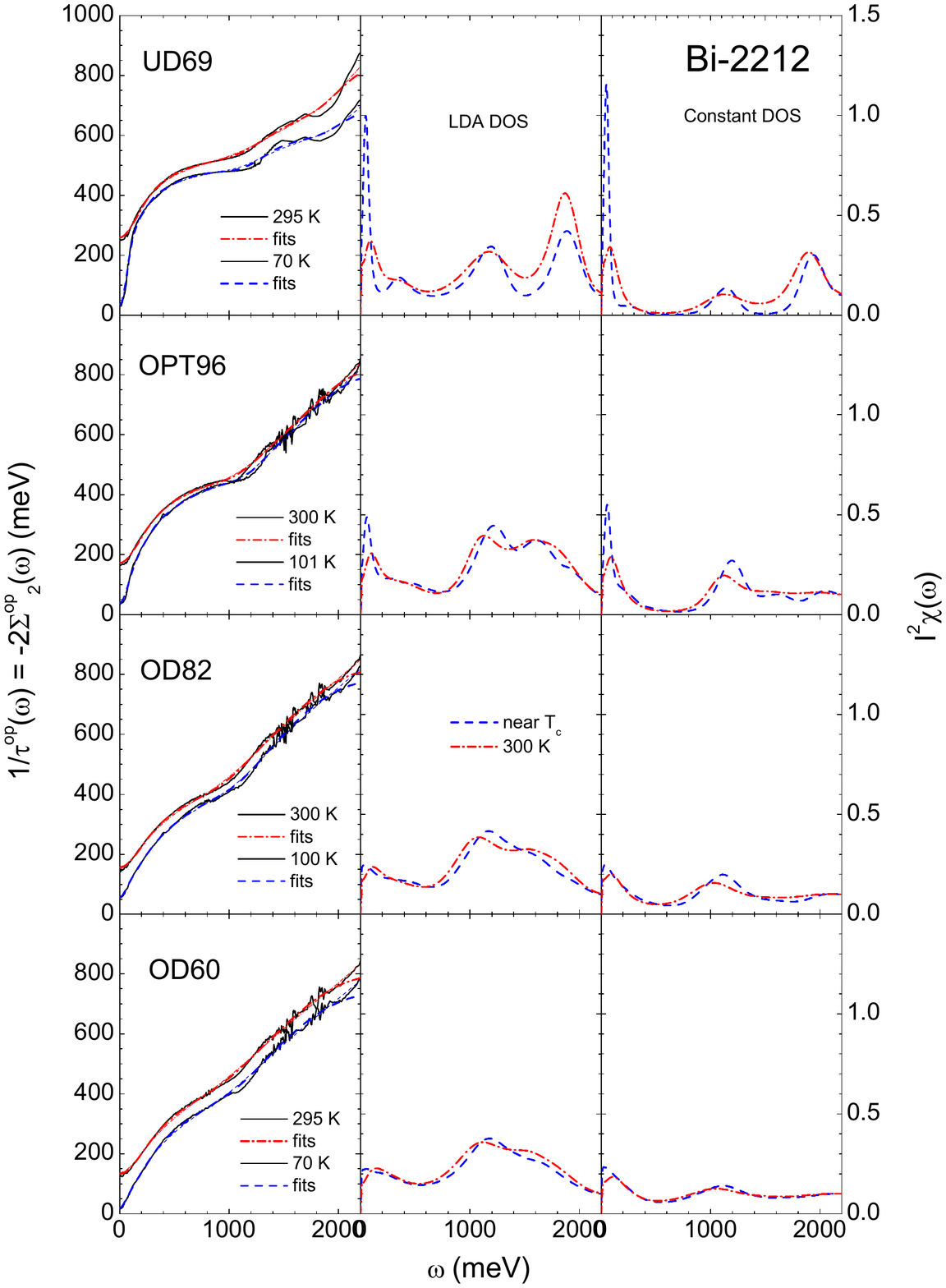}}%
  \vspace*{-0.9 cm}%
\caption{(Color online) The top left frame is the optical scattering rate in meV as a function of $\omega$ in meV for UD69 at two temperatures, next frame below is for OPT96, next OD82 and lower OD60 all for Bi-2212 from reference \cite{hwang:2004,hwang:2007a}. The middle column gives the recovered fluctuation spectrum $I^2\chi(\omega)$ including the DLA density of states in the inversion process while the right column assumes a constant density of states and is for comparison.}
 \label{fig1}
\end{figure}

The equation that needs to be inverted to recover the electron-boson spectral density $I^2\chi(\omega)$, which carries the information on the fluctuation spectrum that scatters the charge carriers has the general form
\begin{equation}\label{eq2}
\frac{1}{\tau^{op}(T,\omega)}=\int d\Omega I^2\chi(T,\Omega) K(T,\omega,\Omega)
\end{equation}
where the kernel $K(T,\omega,\Omega)$ is given in Eqn. (\ref{eq1a}). The deconvolution of this equation to recover an effective spectral density, $I^2\chi(\Omega)$ is ill-conditioned and here we use a maximum entropy technique\cite{schachinger:2006}. The equation can be discretized $D_{in}(i) = \sum_j K(i,j)I^2\chi(j)\Delta\Omega$ where $\Delta\Omega$ is the differential increment on the integration over $\Omega_j = j \Delta\Omega$ with $j$ an integer. We define a $\chi^2$ by
\begin{equation}\label{eq3}
\chi^2 = \sum_{i = 1}^{N}\frac{[D_{in}(i)-\Sigma(i)]^2}{\sigma_i^2}
\end{equation}
where $D_{in}(i)$ is the input data for the optical scattering rate $1/\tau^{op}(i)$ and $\Sigma(i)\equiv\sum_j K(i,j) I^2\chi(j)\Delta\Omega$ is calculated from the known kernel and a given choice of $I^2\chi(j)$, and $\sigma_i$ is the error assigned to the data $D_{in}(i)$. The constraint that the boson exchange function is positive is noted and the entropy functional
\begin{equation}\label{eq4}
L = \frac{\chi^2}{2} -aS
\end{equation}
is minimized with the Shannon-Jones entropy, $S$
\begin{equation}\label{eq5}
S = \int^{\infty}_{0} \Big{[}I^2\chi(\Omega) - m(\Omega) - I^2\chi(\Omega)\ln\Big{|} \frac{I^2\chi(\Omega)}{m(\Omega)}\Big{|} \Big{]}.
\end{equation}
The parameter $a$ in Eqn. (\ref{eq4}) controls how close a fit to the data is obtained. The parameter $m(\Omega)$ is here taken to be some constant value on the assumption that there is no a priori knowledge of the functional form of the electron-boson spectral density $I^2\chi(\Omega)$\cite{schachinger:2006}.

\section{III. Numerical Results}

\begin{figure}[t]
  \vspace*{-1.0 cm}%
  \centerline{\includegraphics[width=3.0 in]{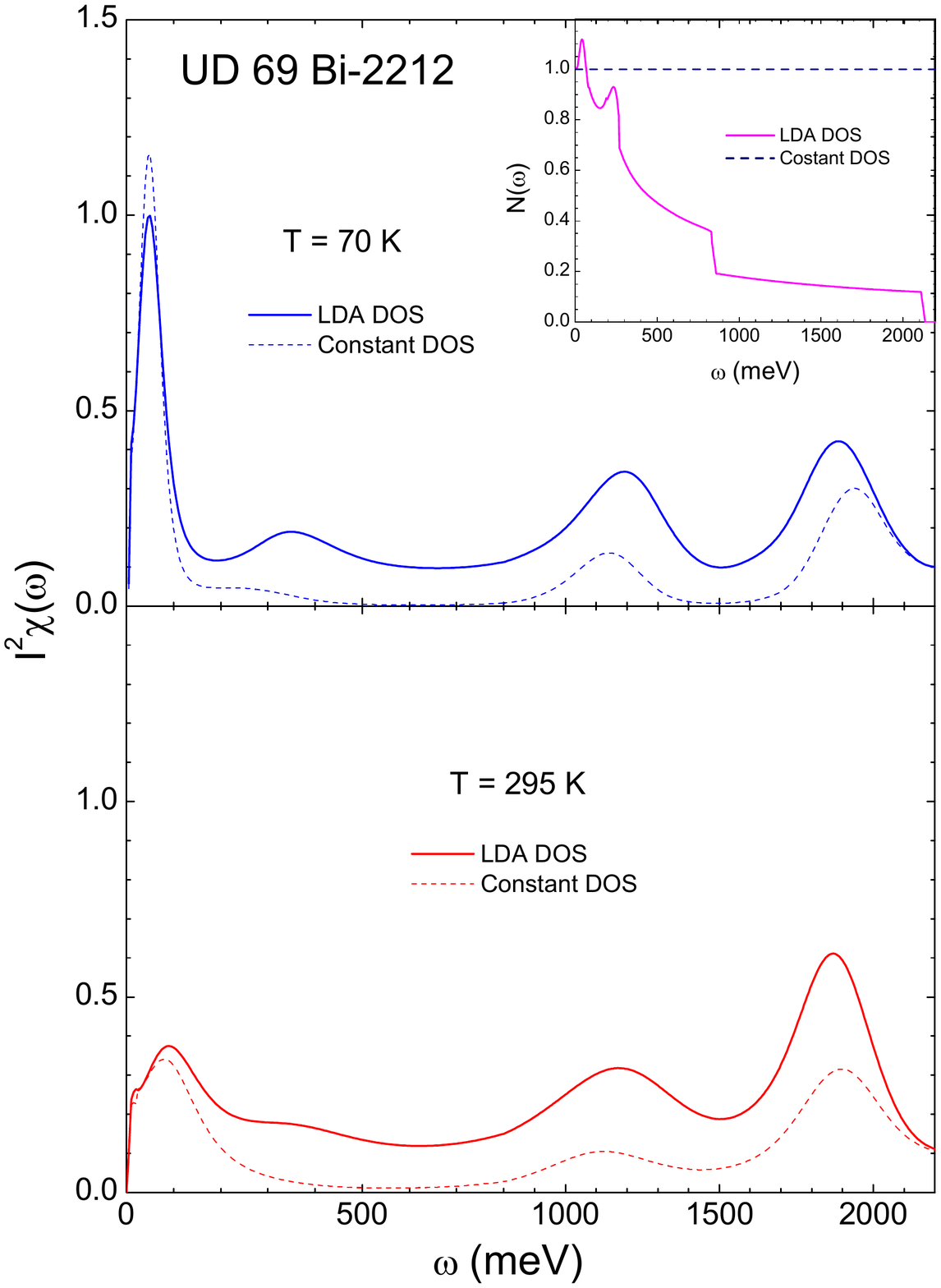}}%
  \vspace*{-0.9 cm}%
\caption{(Color online) The recovered electron-boson spectral density from a maximum entropy inversion of the optical scattering rate for UD69 Bi-2212\cite{hwang:2004,hwang:2007a}. The solid line was obtained accounting for energy dependence in the density of states using LDA data presented in reference\cite{markiewicz:2012} and reproduced here in the inset on the top frame. The dashed curve was obtained assuming a constant density of states and is for comparison with the solid curve. The top frame is for temperature $T =$ 70 K the bottom for $T =$ 295 K.}
 \label{fig2}
\end{figure}

\begin{figure}[t]
  \vspace*{-1.0 cm}%
  \centerline{\includegraphics[width=3.0 in]{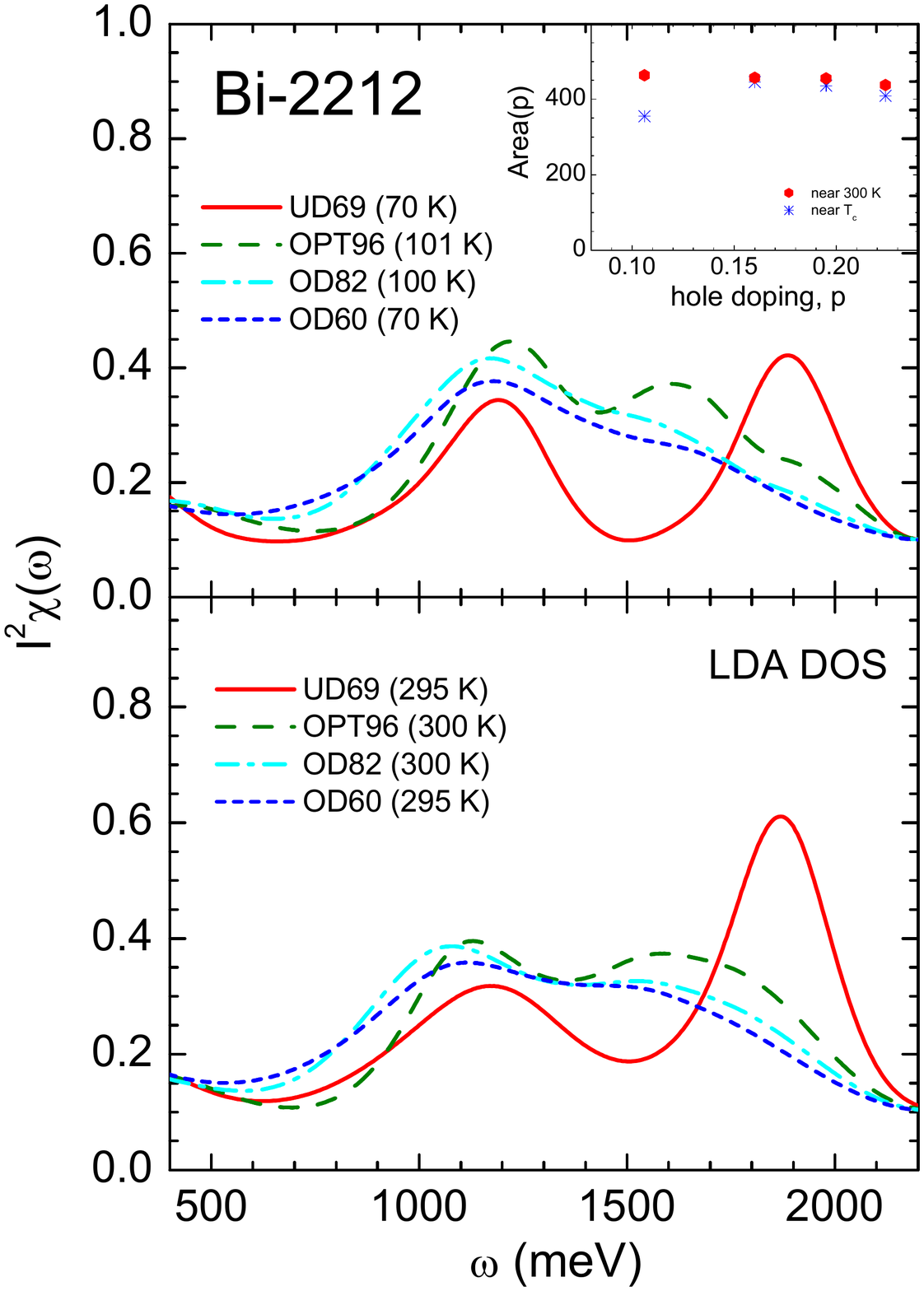}}%
  \vspace*{-0.9 cm}%
\caption{(Color online) The electron-boson spectral density $I^2\chi(\omega)$ as a function of $\omega$ in meV in the range 400 meV to 2200 meV in Bi-2212 for UD60 solid (red), OPT96 dashed(dark green), OD82 dash-dotted (cyan) and OD60 short dashed (blue). Temperatures are as shown in the figure. The LDA density of states shown in the inset of Fig. \ref{fig2} was used. In the inset we show the doping dependent area of $I^2\chi(\omega)$ from 400 meV through 2200 meV.}
 \label{fig3}
\end{figure}

\begin{figure}[t]
  \vspace*{-1.0 cm}%
  \centerline{\includegraphics[width=3.0 in]{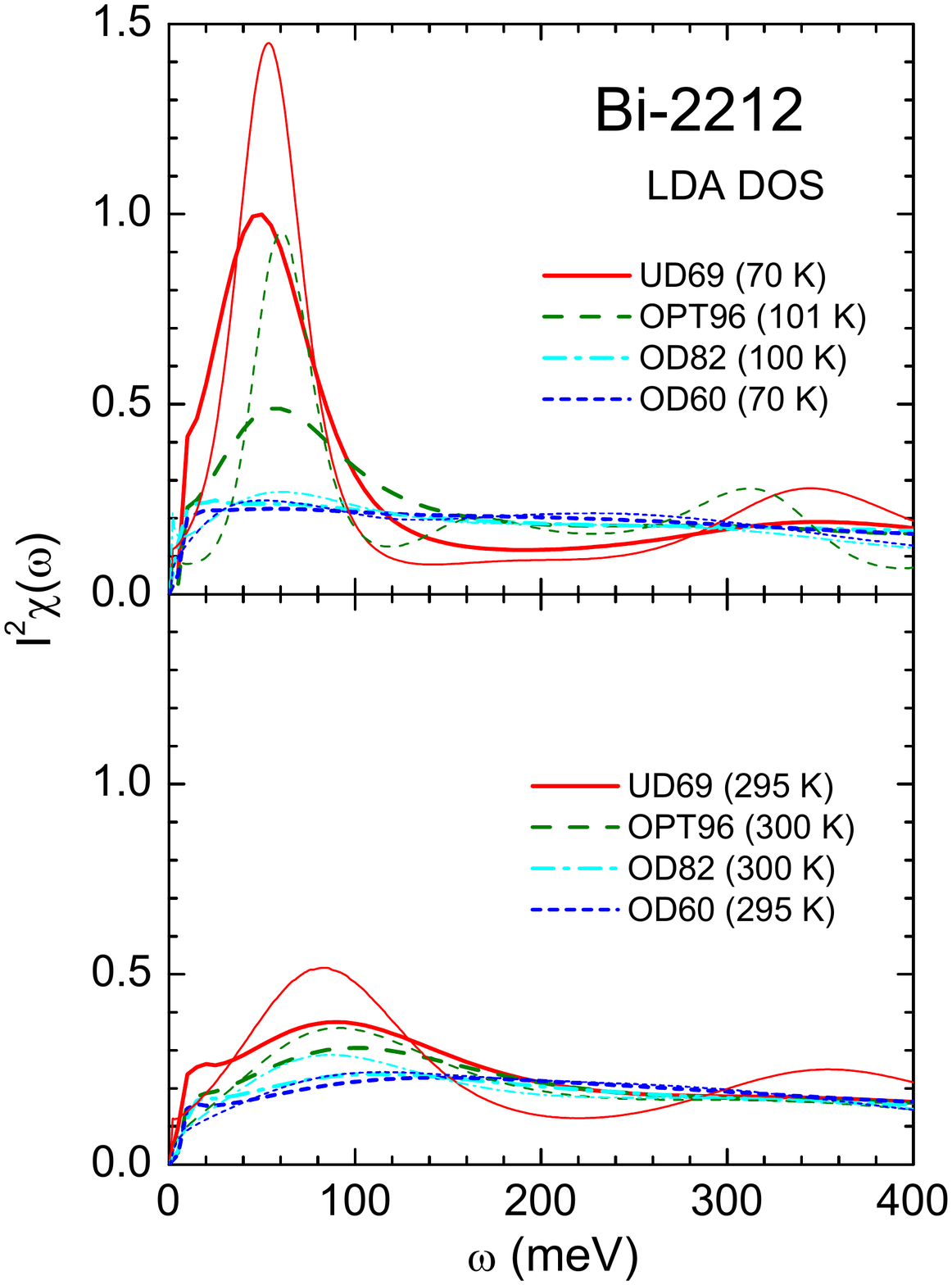}}%
  \vspace*{-0.9 cm}%
\caption{(Color online) The electron-boson spectral density $I^2\chi(\omega)$ as a function of $\omega$ in meV recovered from maximum entropy inversion of the optical scattering rate data in the low energy range to 400 meV in Bi-2212 for UD69 solid (red), OPT96 dashed (dark green), OD82 dash-dotted (cyan) and OD60 short dashed (blue). The temperatures are shown in the figure. The LDA density of states shown in the inset of Fig. \ref{fig2} was used. The heavy curves were obtained when the entire range of the optical data to 2200 meV was used while in the light curves, only data up to 500 meV was employed in the fit.}
 \label{fig4}
\end{figure}

\begin{figure}[t]
  \vspace*{-1.0 cm}%
  \centerline{\includegraphics[width=3.0 in]{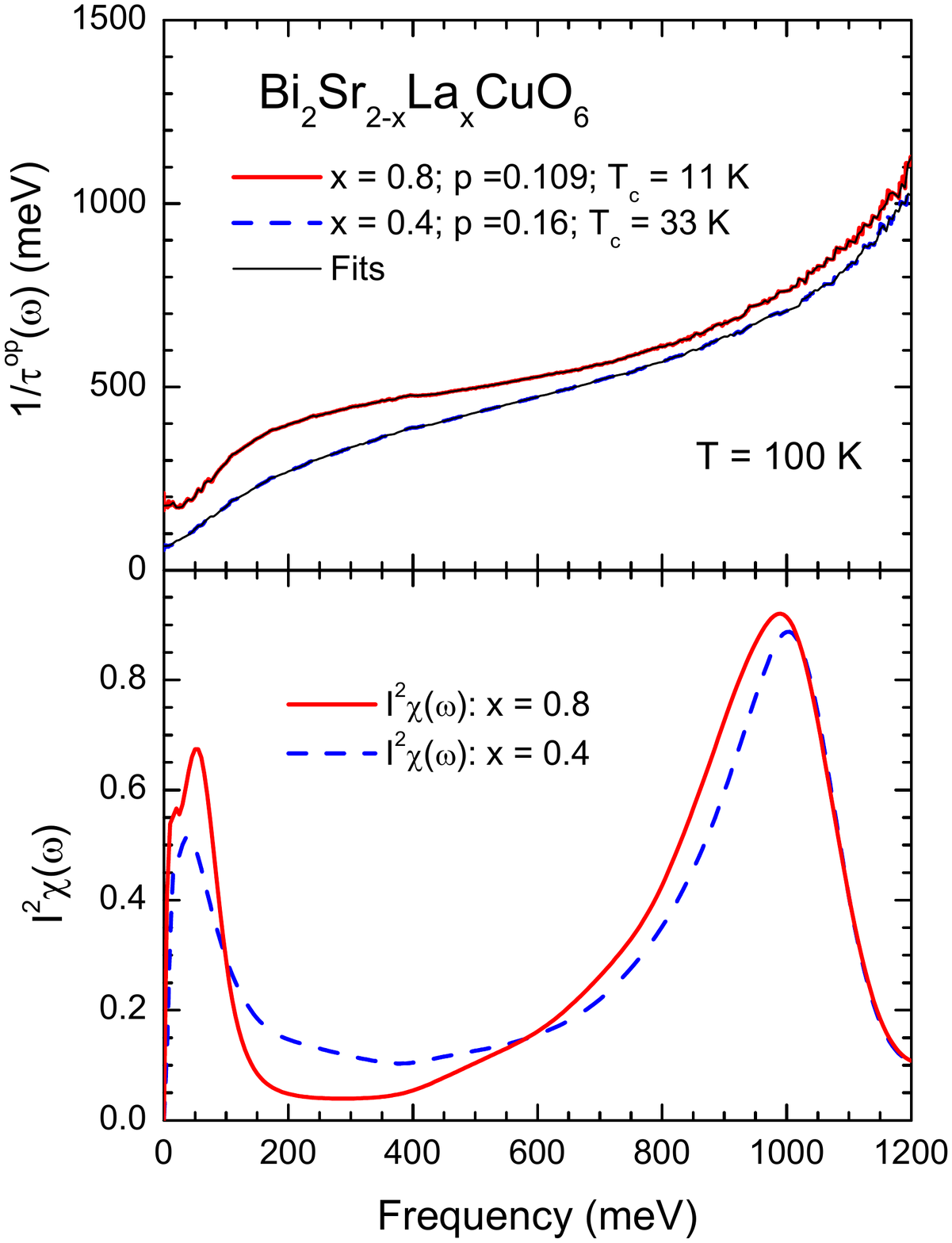}}%
  \vspace*{-0.9 cm}%
\caption{(Color online) The optical scattering rate for two samples of Bi-2201 $x =$ 0.8 solid (red) and $x =$ 0.4 dashed (blue) at $T =$ 100 K from reference\cite{dai:2012} and our maximum entropy fits (upper frame). The recovered electron-boson spectral density $I^2\chi(\omega)$ vs $\omega$ based on the assumption of a constant density of states.}
 \label{fig5}
\end{figure}

In Fig. \ref{fig1} we show results for Bi-2212 based on the data of Hwang, Timusk and Gu\cite{hwang:2004,hwang:2007a}. Four samples representing different doping levels namely UD69, OPT96, OD82, and OD60 are shown from top to bottom. In all cases two temperatures are presented. The first near but above $T_c$ (superconducting critical temperature) (blue) the other near 300 K (red). The continuous lines are the data and dash-dotted our maximum entropy fits. The middle column gives the recovered boson spectral density $I^2\chi(\omega)$ when we use the band structure density of states obtained by LDA which was presented in the work of Markiewicz {\it et al.}\cite{markiewicz:2012} while the right column presents similar data obtained assuming a constant DOS and is for comparison. The particle-hole symmetrized LDA DOS used is reproduced in the inset on the top frame of Fig. \ref{fig2} from reference\cite{markiewicz:2012}. These authors calculate the charge and spin susceptibilities for Bi-2212 and argue that the fluctuation spectrum extends over the full band width of $\geq$ 2.0 eV in this material. Our recovered $I^2\chi(\omega)$ are in general qualitative agreement with this notion. Besides the large peak seen at low energies in UD69 and OPT96 we find that considerable spectral weight resides above 1.0 eV to the highest energy probed 2.2 eV which is the cutoff in the LDA density of states (see inset in Fig. \ref{fig2}). The fluctuation spectra above 1.0 eV is large for all four doping levels considered. It does not vary much with sample and remains at the highest temperature considered. Note that the LDA density of states that we have used in our inversions of the optical data is depressed to 20 \% of its low energy value throughout the region $\omega \geq$ 1.0 eV. This is consistent with the fact that the fluctuation spectra above 1.0 eV is considerably smaller in magnitude in the right frame than in the middle frame. Both columns agree however qualitatively and support the conclusion that $I^2\chi(\omega)$ is finite up to the band cutoff and this conclusion is not importantly dependent on the exact shape of the underlying particle-hole symmetrized DOS. In Fig. \ref{fig2} we make a closer comparison between these two cases for one sample UD69. The results including the LDA density of states are solid curves and the dashed assume the DOS to be constant. The agreement to 100 meV between these two curves is excellent as we might have expected from the form of the LDA DOS which is reasonably approximated by a constant value of one in our units. But this DOS decreases with increasing $\omega$ rather than stay constant so that the fluctuation spectra obtained becomes much larger than in the constant DOS case because the DOS at these energies has been reduced. Nevertheless both spectra show the same overall qualitative behavior. In particular the existence of significant spectral weight in the 1.0 to 2.0 eV region is clearly established even in the dashed curve. While the DOS in this energy region is only 20 \% of its value near $\omega =$ 0, the resulting spectral density is suppressed by much less than a factor of 5.

The region of $I^2\chi(\omega)$ above 400 meV is further emphasized in Fig. \ref{fig3} where results with LDA DOS at different doping are overlaped for easy comparison. There is some evolution of these spectra with doping. For example UD69 and OPT96 clearly have a two peak structure, but these are not separately resolved in other two spectra. On the other hand the overall energy scale involved is not much affected nor is the area under the spectral density even as the temperature is varied. The inset in Fig. \ref{fig3} shows the area (or spectral weight) under the spectral density $I^2\chi(\omega)$ from 400 meV to 2200 meV as a function of doping for the two temperatures investigated. All differences are small and we conclude that these high energy structures are robust and not much dependent on doping level and on temperature. The low energy part of the spectrum is also of interest and many inversions of optical data in the range to 300 meV or so have already appeared in the literature\cite{carbotte:2011,hwang:2007,hwang:2008c}. Here two new items arise. The first is how different is the spectrum when a structured more realistic LDA band density of states such as is seen in the inset of Fig. \ref{fig2} is introduced rather than the constant DOS that has been used in previous inversions\cite{hwang:2007}. We have already commented on this in our discussion of Fig. \ref{fig2}. While the agreement below 100 meV is excellent the magnitude of $I^2\chi(\omega)$ in the region between 100 and 300 meV is much larger in our new inversions (solid) than it was in previous inversions (dashed) and is more consistent with the idea that boson exchange spectral weight remains significant up to high energies (see Fig. \ref{fig2}). The second item is how much effect is there on the recovered $I^2\chi(\omega)$ in the low energy region when one fits data not just up to 300 meV but to 2.2 eV. This is addressed in Fig. \ref{fig4}. Both frames employ the LDA density of states. The heavy solid (red) curve is for UD69, dashed (dark green) for OPT96, dash-dotted (cyan) for OD82 and short dashed (blue) for OD60. These results qualitatively agree with previous results\cite{hwang:2007} based on a constant DOS and data limited to 400 meV. There are however some quantitative differences. Part of the differences is due to our use of the LDA density of states but another part is due to our new fits which include a much larger range of optical data up to 2.2 eV. The light lines color-coded as the heavy line make this comparison. These involve truncating the data at 400 meV. This leads to changes in the height of the first peak in UD69 and OPT96 at $T =$ 70 K and $T =$ 101 K respectively, with more minor shifts in the position of the peaks. As doping is increased into the overdoped region of the Bi-2212 phase diagram the difference between heavy and light curves diminish although some small changes still remain on the scale of our graphs.

In Fig. \ref{fig5} we show additional results for Bi-2201 based on the data of Dai {\it et al.}\cite{dai:2012}. The top frame shows the data on the optical scattering rate $1/\tau^{op}(\omega)$ vs $\omega$ up to 1.2 eV for two samples; we obtained the reflectance spectra from the paper by Dai {\it et al.}\cite{dai:2012} and performed further analysis to get the optical scattering rates. The first (solid red) is $x =$ 0.8 which corresponds to a doping of $p =$ 0.109 and the second (dashed blue) is for $x =$ 0.4 ($p =$ 0.16). Our maximum entropy fits are good and are represented by light continuous curves in the top frame. The recovered spectra for $I^2\chi(\omega)$ are shown in the lower frame. Besides a low energy peak already investigated before\cite{hwang:2013} the spectrum has a large peak centered at $\sim$ 1.0 eV which depends little on doping. These peaks were obtained using a constant density of states approximation and could be even higher since the real electron-hole symmetrized band structure DOS is expected to decrease significantly in magnitude as the band edge is approached. What is clear however is that the fluctuation spectra of Bi-2201 also involves high energy excitations which extend roughly from 600 meV to 1200 meV rather than from 1000 to 2200 meV in the Bi-2212 family.

\section{IV. Summary and Conclusions}

We have obtained the electron-boson spectral density $I^2\chi(\omega)$ up to high energies for Bi-2212 as well as Bi-2201 from data on the optical scattering rate. To extract this information from the available data we used a maximum entropy technique to invert Eqn. (\ref{eq2}) with the kernel in this equation taking full account of the energy dependent density of states determined in LDA calculations for the case of Bi-2212. While all the inversions done so far (that we are aware of) have limited the energy ranged used to 300 meV or so, here we have gone to much large energies up to 2.2 eV to investigate the possibility that the fluctuation spectra may extend in these materials up to the band edge. To do this it was essential to generalize existing inversion techniques to include variation in the DOS over such a large energy scale. We find that peaks exist in $I^2\chi(\omega)$ in the region of 1.0 eV to 2.2 eV for Bi-2212 and for Bi-2201 in the region 600 to 1200 meV. Such high energy excitations with significant spectral weight appear to be a general feature of the cuprates.

%
%
\acknowledgments JH acknowledges financial support from the National Research Foundation of Korea (NRFK Grant No. 20100008552). JPC was supported by the Natural Science and Engineering Research Council of Canada (NSERC) and the Canadian Institute for Advanced Research (CIFAR).

%
%

\bibliographystyle{apsrev4-1}
\bibliography{bib}

\end{document}